\documentclass{iau} 
\usepackage{graphicx}	
\usepackage{amsmath}	
\usepackage{amssymb}	
\usepackage[para,flushleft]{threeparttable} 
\usepackage{epstopdf}
\usepackage{textgreek}
\usepackage{comment}
\usepackage{hyperref}

\usepackage{xcolor, fontawesome}
\definecolor{twitterblue}{RGB}{64,153,255}
\newcommand{\twitter}[1]{\href{https://twitter.com/#1}{\textcolor{twitterblue}{\faTwitter}\,\tt\hspace{2pt}\textcolor{blue!60!black}{@#1}}}

\hypersetup{colorlinks,linkcolor={red!60!black},citecolor={blue!60!black},urlcolor={blue!60!black}}

\usepackage{natbib,twoopt}
\bibpunct{(}{)}{;}{a}{}{,} 
\makeatletter
\newcommandtwoopt{\citeads}[3][][]{\href{http://ui.adsabs.harvard.edu/\#abs/#3}%
{\def\hyper@linkstart##1##2{}%
\let\hyper@linkend\@empty\citealp[#1][#2]{#3}}}
\newcommandtwoopt{\citepads}[3][][]{\href{http://ui.adsabs.harvard.edu/\#abs/#3}%
{\def\hyper@linkstart##1##2{}%
\let\hyper@linkend\@empty\citep[#1][#2]{#3}}}
\newcommandtwoopt{\citetads}[3][][]{\href{http://ui.adsabs.harvard.edu/\#abs/#3}%
{\def\hyper@linkstart##1##2{}%
\let\hyper@linkend\@empty\citet[#1][#2]{#3}}}
\newcommandtwoopt{\citeyearads}[3][][]%
{\href{http://ui.adsabs.harvard.edu/\#abs/#3}
{\def\hyper@linkstart##1##2{}%
\let\hyper@linkend\@empty\citeyear[#1][#2]{#3}}}
\makeatother

\title[(Simulating) CMEs in Active Stars] 
{(Simulating) Coronal Mass Ejections in Active Stars}

\author[J.~D.~Alvarado-G\'omez~et~al.]   
{Juli\'an D. Alvarado-G\'omez$^{1,2,\thanks{Karl Schwarzschild Fellow}}$, Jeremy~J.~Drake$^{2}$, Cecilia Garraffo$^3$, Sofia~P.~Moschou$^2$, Ofer Cohen$^4$, Rakesh K. Yadav$^3$ \and \hspace{10cm} Federico Fraschetti$^{2,5}$}

\affiliation{$^1$Leibniz Institute for Astrophysics Potsdam \\ An der Sternwarte 16, 14482 Potsdam, Germany \\ email: {\tt julian.alvarado-gomez@aip.de} $|$ \twitter{AstroRaikoh}\\
[\affilskip]
$^2$Center for Astrophysics $|$ Harvard \& Smithsonian\\ 60 Garden Street, Cambridge, MA 02138, USA\\
[\affilskip]
$^3$Institute for Applied Computational Science, Harvard University\\ 33 Oxford Street, Cambridge, MA 02138, USA\\
[\affilskip]
$^4$University of Massachusetts at Lowell, Department of Physics \& Applied Physics\\ 600 Suffolk Street, Lowell, MA 01854, USA\\
[\affilskip]
$^5$Department of Planetary Sciences-Lunar and Planetary Laboratory\\ University of Arizona, Tucson, AZ 85721, USA\\}

\pubyear{2020}
\volume{354}  
\jname{Solar and Stellar Magnetic Fields: Origins and Manifestations}
\editors{A. Kosovichev, K. Strassmeier \& M. Jardine, eds.}
\begin{document}

\maketitle

\begin{abstract}
The stellar magnetic field completely dominates the environment around late-type stars. It is responsible for driving the coronal high-energy radiation (e.g. EUV/X-rays), the development of stellar winds, and the generation transient events such as flares and coronal mass ejections (CMEs). While progress has been made for the first two processes, our understanding of the eruptive behavior in late-type stars is still very limited. One example of this is the fact that despite the frequent and highly energetic flaring observed in active stars, direct evidence for stellar CMEs is almost non-existent. Here we discuss realistic 3D simulations of stellar CMEs, analyzing their resulting properties in contrast with solar eruptions, and use them to provide a common framework to interpret the available stellar observations. Additionally, we present results from the first 3D CME simulations in M-dwarf stars, with emphasis on possible observable signatures imprinted in the stellar corona.  
\end{abstract}

\firstsection 
\section{Introduction}

\noindent Flares and Coronal Mass Ejections (CMEs) are spectacular manifestations of magnetic energy release in the Sun and cool stars. Flares correspond to a temporal increase in the electromagnetic radiation (across the entire spectrum) up to several orders of magnitude, in rare cases briefly exceeding a star's quiescent state bolometric luminosity (in particular wavelengths). A CME is characterized by the release of relatively dense, magnetized material to the outer corona and the stellar wind (\citeads{2012LRSP....9....3W}, \citeads{2017LRSP...14....2B}). Large flares on the Sun (X-class, or $\ge 10^{31}$ ergs in 1--8 \AA \,\,soft X-rays) are almost always associated with CMEs (e.g, \citeads{2009IAUS..257..233Y}). The most energetic events pose threats to life on Earth, with recent examples including the X15 and X4 class flares (nearly $10^{33}$ erg of X-ray energy) of March 6 1989 that triggered a CME that caused the collapse of Quebec's electricity grid; and the 1859 ``Carrington Event'' that ignited telegraph lines and spread aurorae as far south as Hawaii, Cuba, and even Colombia (\citeads{1859MNRAS..20...13C}, \citeads{2016AdSpR..57..257M}). An extraordinary event in AD 774-775, that was discovered in $^{14}$C/$^{12}$C data from Japanese cedar tree rings, was also likely caused by a huge solar flare that would have destroyed 20\% of the ozone layer \citepads{2012Natur.491E...1M}.

Despite their importance, the energetics of flares and CMEs is still poorly constrained. Flares originate with the sudden conversion of magnetic energy into plasma heating and acceleration of electrons and protons within a magnetic loop. Observations have revealed that the radiated energy in white light completely dominates the soft X-ray emission in solar flares\,---by factors of up to 100 (\citeads{2011A&A...530A..84K}, \citeads{2012ApJ...759...71E}). In turn, the kinetic energy of the CMEs associated with large flares are typically larger than the bolometric luminosity by factors of 3 or so. Moreover, multi-spacecraft observations have shown that only up to $20$\% of the kinetic energy of CMEs is channelled into generation of energetic particles \citepads{2012ApJ...759...71E}. The solar data, then, indicate that soft X-rays constitute only a few percent of the total dissipated energy, and that CMEs carry more energy than flares.

The situation in the stellar regime may be radically different. Large flares on active M-dwarfs and T Tauri stars can reach total soft X-ray fluences of $10^{34} - 10^{36}$ erg\,---between three and five orders of magnitude larger than in X-class solar flares (e.g. \citeads{2004A&A...416..713G}, \citeads{2019A&A...622A.210G}). Furthermore, the coronae of very active stars appear to be continuously flaring (e.g. \citeads{1997ApJ...480L.121G}, \citeads{2000ApJ...545.1074D}, \citeads{2010ApJ...723.1558H}), and their light curves can be well-modeled using a superposition of flares (e.g.~\citeads{2002ApJ...580.1118K}, \citeads{2007A&A...471..645C}). This presents a serious energy problem. The most active stars are observed to emit about 1/1000th of their bolometric luminosity in soft X-rays\,---the so-called coronal saturation limit (e.g.~\citeads{2011ApJ...743...48W}, \citeads{2018MNRAS.479.2351W}). If these X-rays originate from flares, and only about 1\% of the flare energy is in the form of soft X-rays like in solar flares, then the implication is that magnetic energy dissipation\,---mainly in the form of CMEs---\,amounts to 100 times the X-ray flux or about 10\% of a star's total energy output \citepads{2013ApJ...764..170D}. Apart from placing an implausibly high energy requirement, such elevated stellar CME activity largely disagrees with observations (or the lack of, see~\citeads{2014MNRAS.443..898L}, \citeads{2016ApJ...830...24C}, \citeads{2017PhDT.........8V}, \citeads{2018ApJ...856...39C}), where only one event has been confirmed so far \citepads{2019NatAs...3..742A}. It appears then that strong flares and CMEs on very active stars have a quite different energy partition compared to their solar counterparts. 

Briefly discussed by \citetads{2016IAUS..320..196D}, this re-distribution of energy could be related with a suppression mechanism of CMEs, in which the stellar large-scale magnetic field would entrap the plasma ejecta (up to a certain energy), allowing only the radiation and a fraction of particles accelerated at the flare site to escape (see also \citeads{2017MNRAS.472..876O}, \citeads{2019ApJ...874...21F}). We investigate this possibility through 3D magnetohydrodynamic (MHD) simulations, applying realistic models currently used in space weather studies of the solar system (e.g.~\citeads{2017ApJ...834..172J}). The CMEs evolve in the corona and stellar wind conditions imposed by surface field configurations\,--- in terms of field strength and topology---\, compatible with observations of young Sun-like stars (e.g.~\citeads{2009ARA&A..47..333D}, \citeads{2015A&A...582A..38A}) and a state-of-the-art dynamo simulation of the fully-convective M-dwarf Proxima Centauri (see \citeads{2016ApJ...833L..28Y}). We provide here a summary of the results and refer the reader to Alvarado-G\'omez~et~al. (\citeyearads{2018ApJ...862...93A}, \citeyearads{2019ApJ...884L..13A}) for additional details.  

\firstsection
\section{Numerical Models}

\noindent Three different models are considered in our investigation. The first one is the Alfv\'en Wave Solar Model (AWSoM, \citeads{2013ApJ...764...23S}, \citeads{2014ApJ...782...81V}), which is used to compute the quiescent conditions of the stellar wind and corona (steady-state solution), driven by the magnetic field configuration at the stellar surface. Coupled to AWSoM, the flux-rope models of \citetads[GL,]{1998ApJ...493..460G} and \citetads[TD,]{1999A&A...351..707T} serve to drive the CME simulations. These models are part of the Space Weather Modeling Framework (\citeads[SWMF,]{2018LRSP...15....4G}), commonly used for solar system research and forecast (e.g. \citeads{2008ApJ...684.1448M}, \citeads{2013ApJ...773...50J}, \citeads{2017ApJ...834..173J}, \citeads{2017ApJ...845...98O}). 

\firstsection
\section{Results and Discussion}

\subsection{Suppression of CMEs by a large-scale magnetic field}

We begin by considering a surface field configuration suitable for testing the large-scale magnetic confinement of stellar CMEs. For this, we use the well-studied synoptic magnetogram of the solar Carrington Rotation (CR) 2107 (e.g. \citeads{2013ApJ...764...23S}, Jin~et~al. \citeyearads{2017ApJ...834..172J}, \citeyearads{2017ApJ...834..173J}), and enhance the dipole component (aligned with the stellar rotation axis) to $75$~G\footnote[2]{Restricted by computational limitations.}. While much stronger large-scale fields are reported for very active stars (up to kG levels, see \citeads{2011IAUS..271...23D}), this assumption is commensurable with observed surface magnetic fields in $\sim\,0.4 - 0.8$ Gyr old F-G-K main sequence stars (e.g. \citeads{2012A&A...540A.138M}, \citeads{2014A&A...569A..79J}, \citeads{2016A&A...585A..77H}). Nominal solar values for mass, radius, and rotation period are assumed in our simulations.

\begin{figure*}[!ht]
\centering
\includegraphics[width=0.498\textwidth]{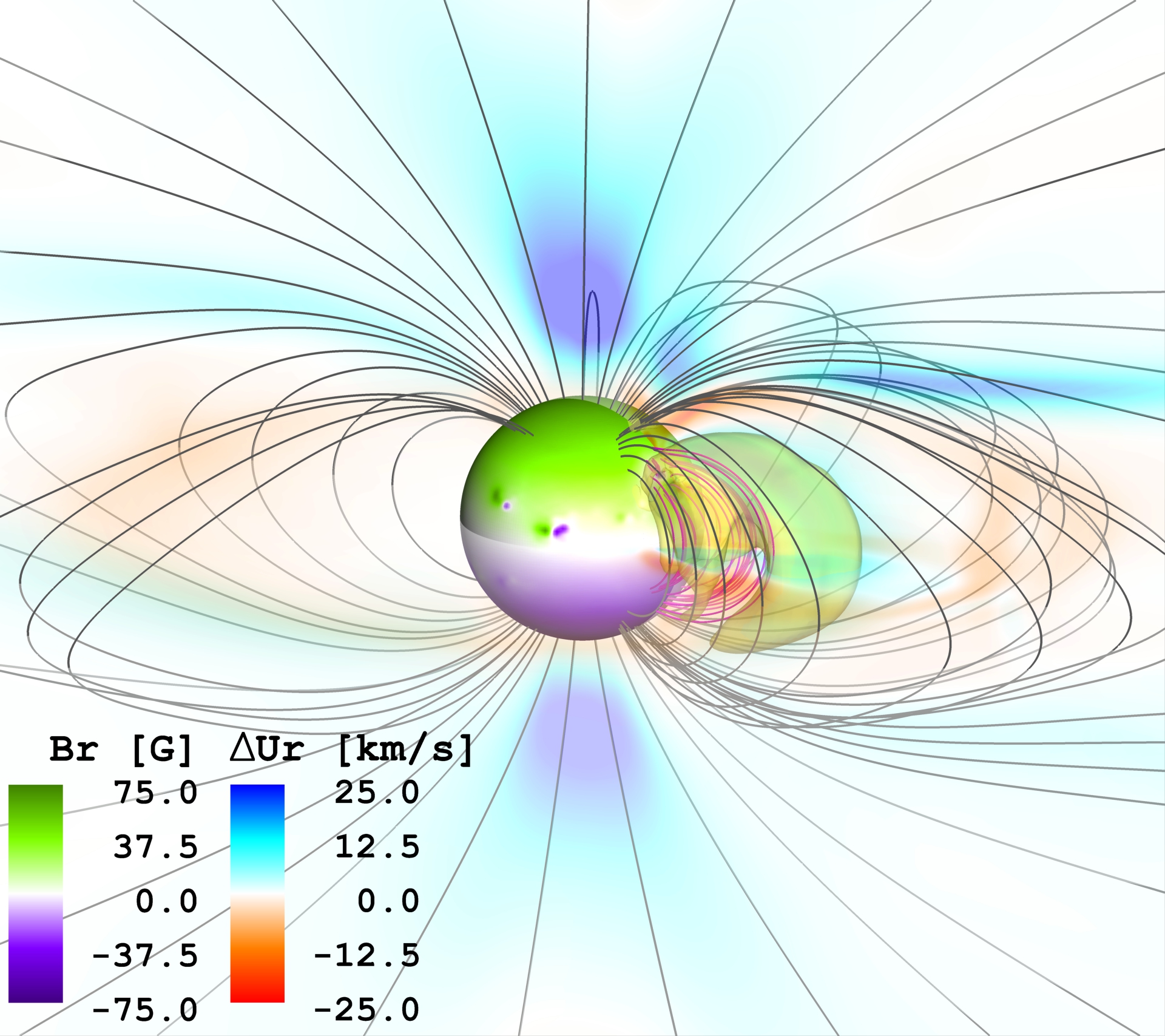}\hspace{1.2pt}\includegraphics[width=0.498\textwidth]{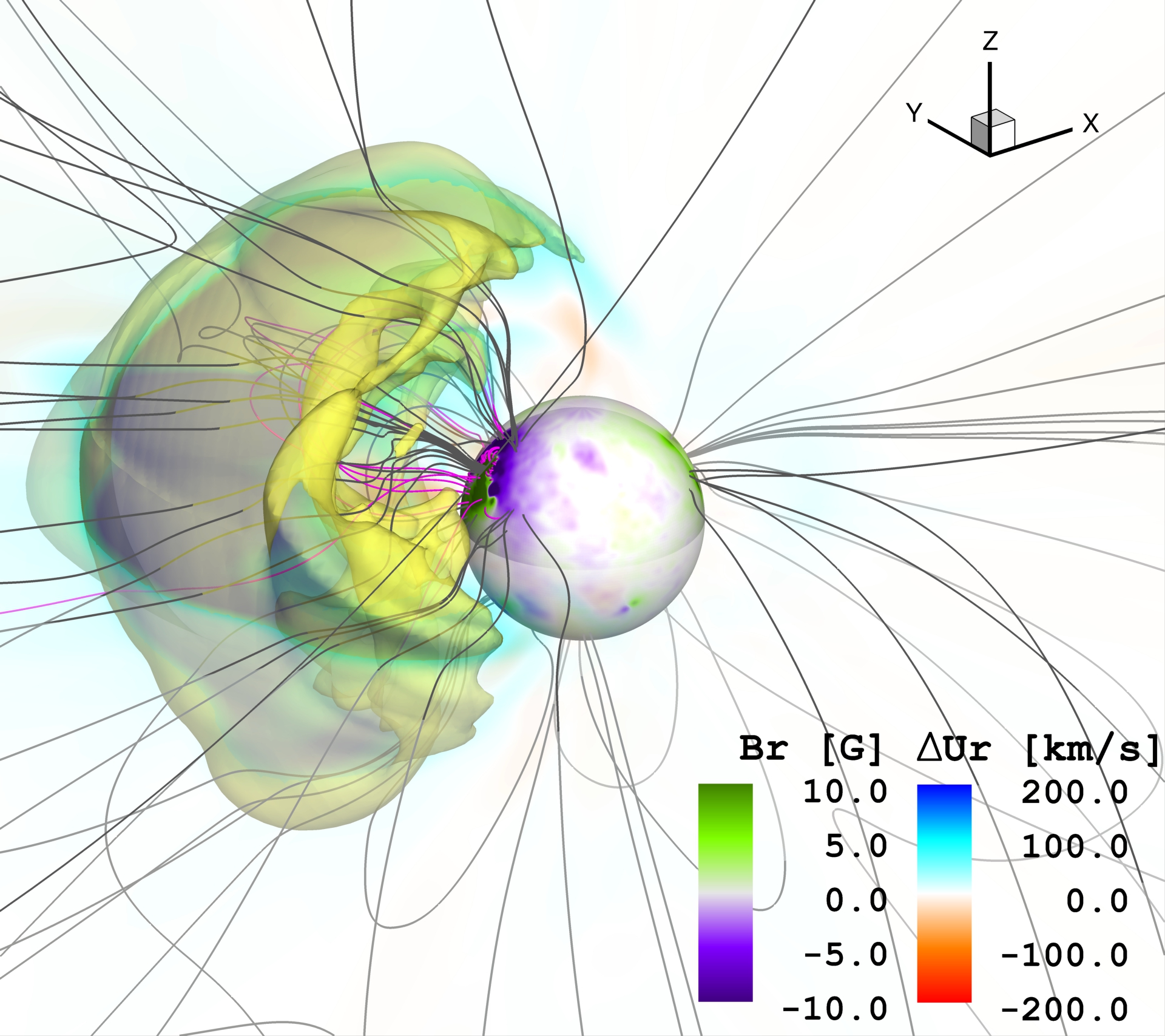}
\caption{Results for a GL flux-rope eruption taking place in AWSoM simulations driven by the CR\,2107\,+\,$75$~G large-scale dipole case (left), and the nominal CR\,2107 (right). The central sphere corresponds to the stellar surface, colored by the magnetic field driving the model. The secondary color scale denotes the coronal Doppler shift velocity ($\Delta U_{\rm r}$) with respect to the pre-CME conditions. The identified eruption is visualized as a translucent yellow iso-surface. Selected magnetic field lines surrounding the eruptive active region (magenta) and associated with the large-scale field (gray) are included. The field of view in both panels is 10~$R_{\bigstar}$.}
\label{fig_1}
\end{figure*}

In addition, we take advantage of the solar CME numerical calibration study (based on the GL flux-rope model) performed by \citetads{2017ApJ...834..172J}, and employ the same eruption parameters in our modified stellar simulations. Figure~\ref{fig_1} shows the same flux-rope eruption taking place under the modified CR\,2107\,+\,$75$~G large-scale dipole field (left) and the fiducial solar CR\,2107 magnetic configuration (right). While this particular eruption, with an associated poloidal flux $\Phi_{\rm p} \simeq 2.0 \times 10^{22}$~Mx (or an equivalent X5.0 GOES class; see \citeads{2018ApJ...862...93A}), produces a relatively strong CME in the solar case ($M^{\rm CME} \sim 10^{17}$~g, $E_{\rm K}^{\rm CME}~\sim~10^{32}$~erg), is totally confined in the stellar simulation. The perturbed material (yellow iso-surface in the visualization) follows the overlying field, remaining bound to the lower regions of the corona. As discussed in \citetads{2018ApJ...862...93A}, we found that eruptive events with equivalent flare energies up to $\sim$X20 in the GOES classification, would be mitigated by this particular configuration of the large-scale magnetic field. 

We also considered sufficiently strong flux-rope eruptions so they would escape the large-scale field confinement. Our analysis revealed that the overlying field significantly reduced the final CME speeds (and therefore the associated kinetic energies) in contrast with extrapolations from solar data. On the other hand, the total mass perturbed in our simulated events (confined and escaping) roughly followed the solar flare-CME relation extended to the stellar regime (\citeads{2012ApJ...760....9A}, \citeads{2013ApJ...764..170D}). Interestingly, these two predictions are consistent with the observed properties of the best stellar CME candidates observed so far (Moschou~et~al.~\citeyearads{2017ApJ...850..191M}, \citeyearads{2019ApJ...877..105M}, \citeads{2019A&A...623A..49V}), as well as with the only direct detection currently available \citepads{2019NatAs...3..742A}.

\subsection{Magnetically-suppressed CME events in M-dwarf stars: Coronal response}

\begin{figure*}[!ht]
\centering
\includegraphics[height=0.469\textwidth]{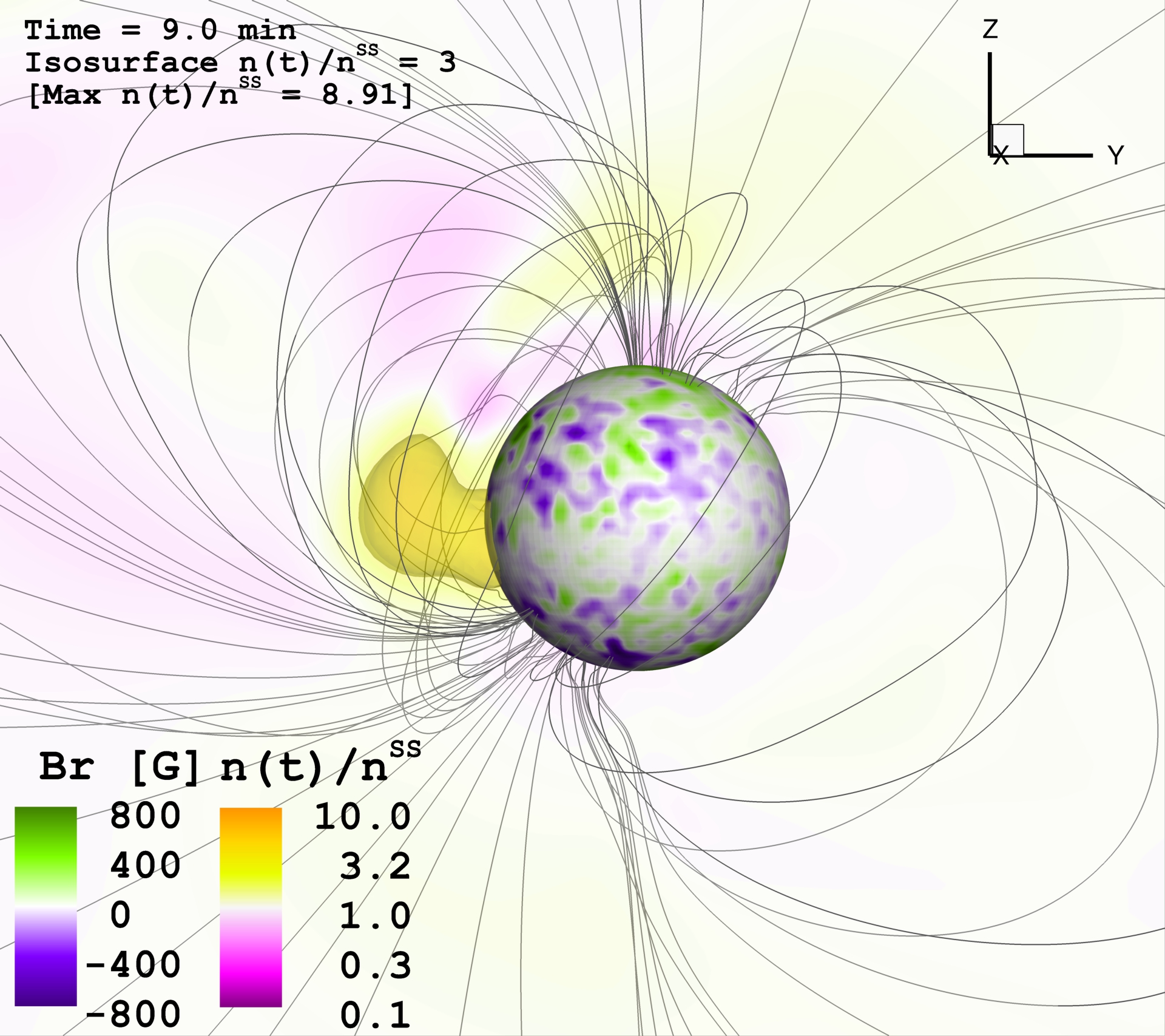}\hspace{1.2pt}\includegraphics[height=0.469\textwidth]{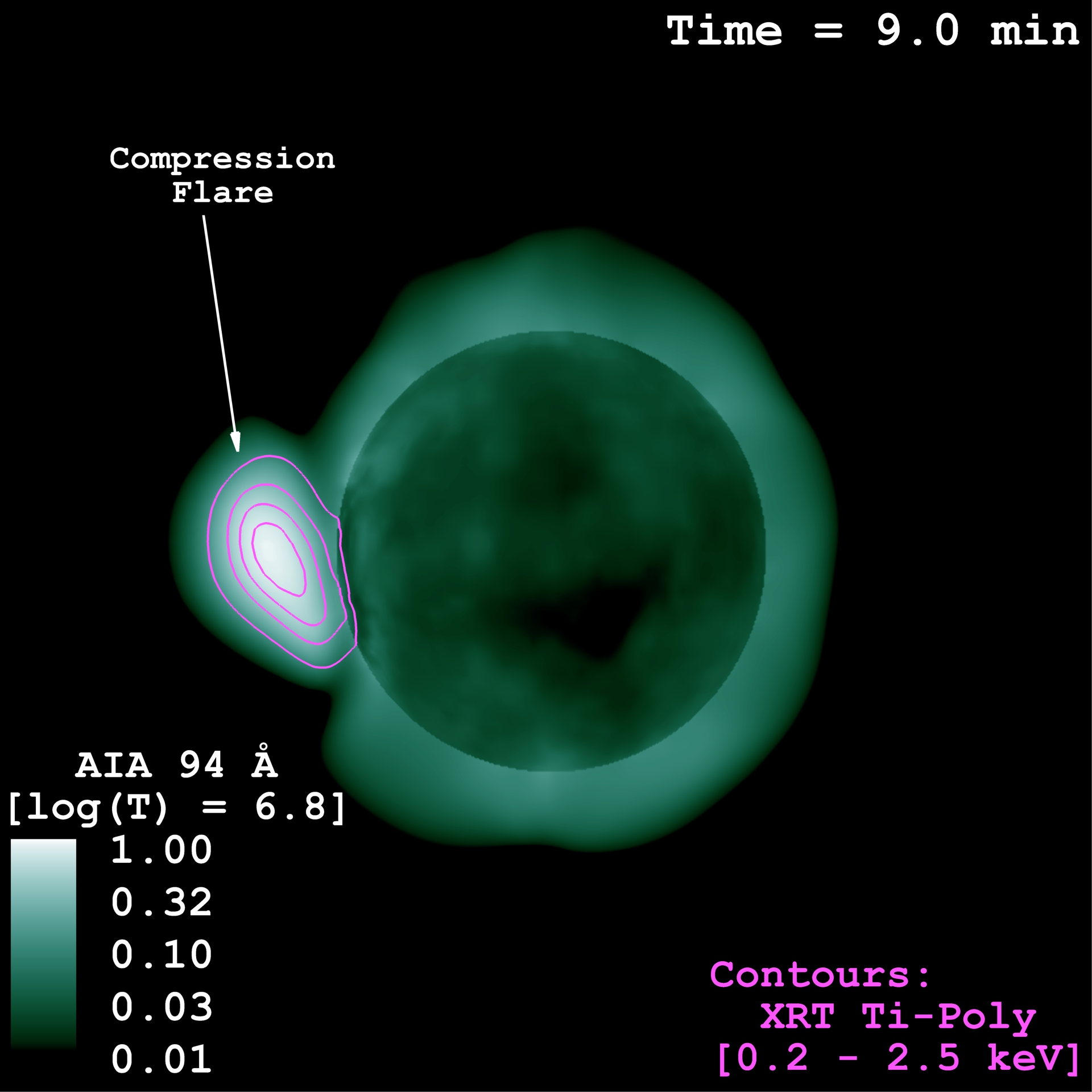}

\vspace{-0.5pt}\includegraphics[height=0.469\textwidth]{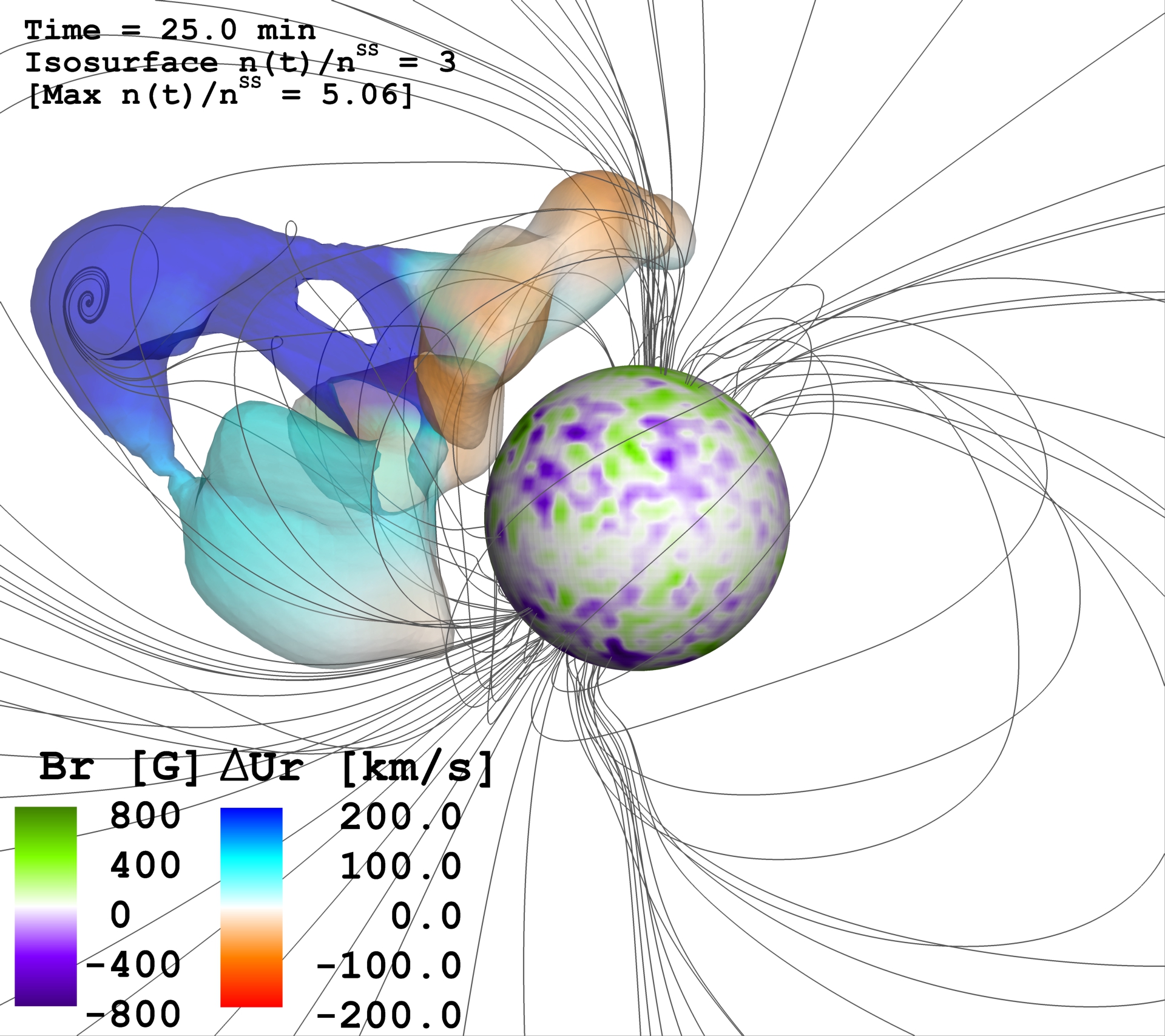}\hspace{1.2pt}\includegraphics[height=0.469\textwidth]{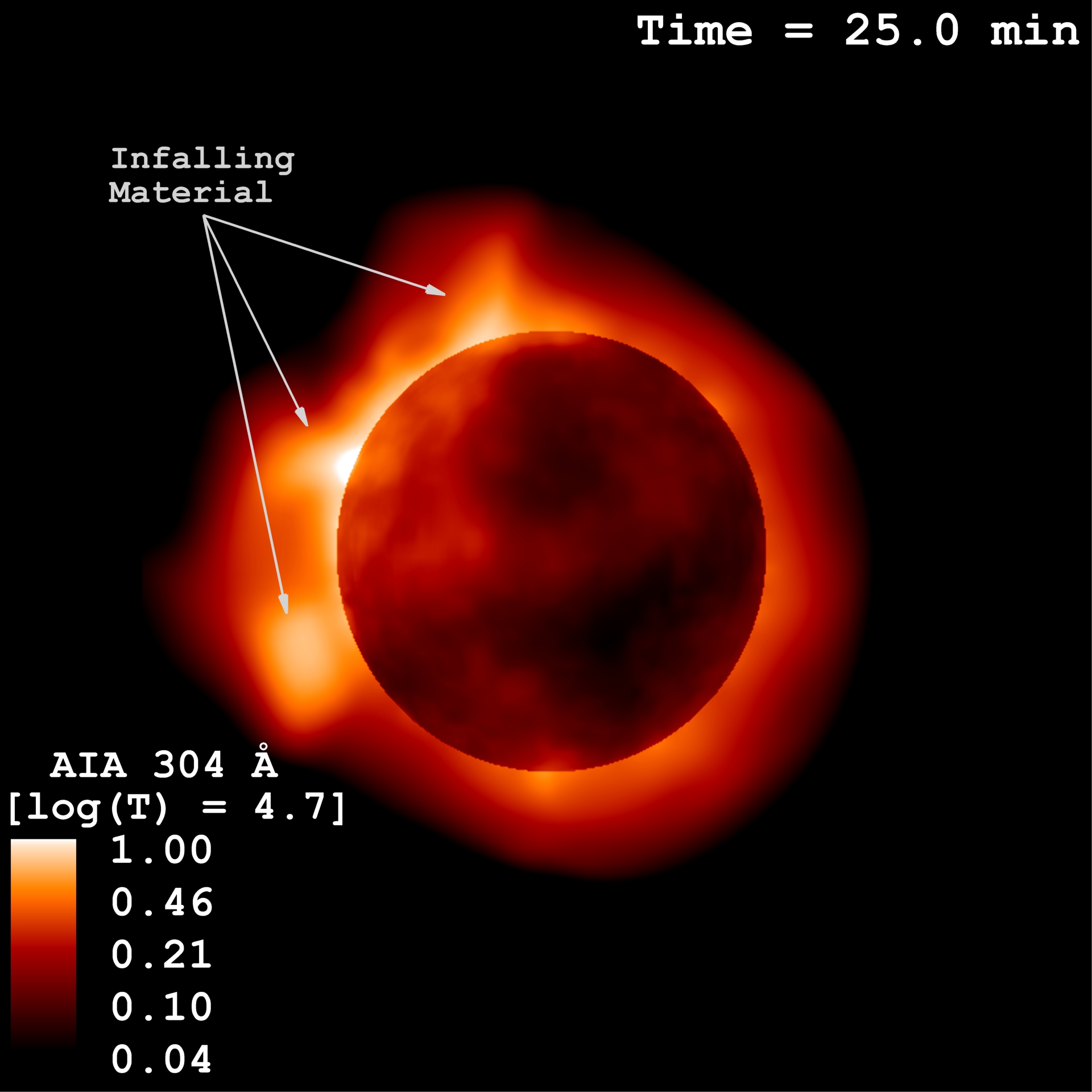}
\caption{Snapshots at different times of a CME simulation in the flare star Proxima Centauri. The magnetic geometry driving the AWSoM corona/stellar wind solution is provided by the fully-convective dynamo simulation of \citetads{2016ApJ...833L..28Y}. The case shown corresponds to a scaling in the surface field between $\pm\,800$~G. \textit{Left:} Three-dimensional visualizations of the density contrast ($n(t)/n^{\rm SS}$, top), and the coronal Doppler shift velocity ($\Delta U_{\rm r}$, bottom). The listed isosurface value is used to identify the perturbation. Selected large-scale magnetic field lines are shown in gray. \textit{Right:} Simulated line-of-sight images of the stellar corona (in arbitrary normalized units) synthesized in two AIA/SDO filters (94~\AA, top; 304~\AA, bottom). Magenta contours in the top panel localize the X-ray emission (Ti-poly filter of XRT/Hinode; $0.2-2.5$~keV) at the peak of the CME-induced compression flare. Arrows in the bottom panel indicate bight kernels of infalling material due to the partial confinement of the CME by the large-scale field.}
\label{fig_2}
\end{figure*}

Similar to the Sun-like models, our M-dwarf CME simulations consist of a steady-state description of the stellar wind and the corona, which is then used as initial condition for a time-dependent flux-rope eruption model (the TD model was considered in this case). We drive AWSoM using the surface field configuration predicted by a self-consistent dynamo simulation of a fully-convective star \citepads{2016ApJ...833L..28Y}. Fine-tuned to the archetypical flare star Proxima Centauri\,---in terms of mass, radius, and rotation period---\, this dynamo model yields a long-term variability time-scale for the stellar magnetic field compatible with the observed activity cycle in this star ($P_{\rm cyc} \sim 7$ yr, \citeads{2016A&A...595A..12S}, \citeads{2017MNRAS.464.3281W}).

As discussed in \citetads{2019ApJ...884L..13A}, we focused on the coronal response during a CME, using eruption parameters expected on the stellar regime ($M^{\rm FR}~=~4~\times~10^{14}$~g, $E_{\rm B, free}^{\rm FR}~\simeq~6.5 \times 10^{34}$~erg), and under different large-scale confinement conditions (weak, moderate, strong). The latter was achieved by scaling the strength of the  dynamo-generated surface field to match values observed in low- to moderately-active M-dwarfs \citepads{2014IAUS..302..156R}. We analyzed five cases, with surface magnetic field strengths ranging from $600$~G to $1400$~G in $200$~G increments. Visualizations of the $\pm\,800$~G scaling case are presented in Fig.~\ref{fig_2}. The main results of our analysis can be summarized as follows:\hspace{1cm}

\smallskip
\begin{itemize}
\item Weakly and partially confined CMEs generated a flare-like signature in the corona, predominantly in the integrated X-ray emission ($0.2 - 2.5$ keV), with increments of up to one order of magnitude with respect to the steady-state pre-CME conditions (e.g.~Fig.~\ref{fig_2}, top-right). While resembling a normal stellar flare, with durations between tens of minutes up to an hour, these brightenings are not powered by magnetic reconnection but instead by strong compression of the coronal material by the escaping CME. The relative strength of the peak in these transient signatures decreases as the CME suppression increases. 

\smallskip
\item The resulting flare-like events show a characteristic hot-to-cool, red-to-blue evolution in their Doppler shift profiles, transitioning from hotter ($\log(T) \gtrsim 6.8$) to cooler ($\log(T) \lesssim 6.0$) coronal lines, and with velocities within $\pm\,200$ km s$^{-1}$ (for the CME and stellar parameters here considered). As with the general behavior of stellar CMEs, stronger confinement leads to lower relative velocities ($<$$100$ km s$^{-1}$).

\smallskip
\item A gradual brightening of the soft X-ray corona (by factors of $\sim2-3$) is observed in fully suppressed CME events. Extending over the course of several hours, the associated emission is redshifted ($<$$-50$~km~s$^{-1}$), indicative of infalling material, which we designate as a \textit{coronal rain cloud}. A similar infalling process, with corresponding brightenings in the lower layers of the corona, is obtained for partially confined events where CME fragmentation takes place (see Fig.~\ref{fig_2}, bottom-right).
\end{itemize}

\acknowledgments
\noindent J.D.A.G. would like to thank the organizers of the IAU Symposium 354 for the invitation to present this work and for the financial support received to attend the conference. J.D.A.G. was also supported by Chandra GO5-16021X and HST GO-15326 grants. J.J.D. was funded by NASA contract NAS8-03060 to the Chandra X-ray Center and thanks the director, Belinda Wilkes, for continuing advice and support. S.P.M. and O.C. were supported by NASA Living with a Star grant number NNX16AC11G. This work was carried out using the SWMF/BATSRUS tools developed at The University of Michigan Center for Space Environment Modeling (CSEM) and made available through the NASA Community Coordinated Modeling Center (CCMC). We acknowledge the support by the DFG Cluster of Excellence "Origin and Structure of the Universe". Some simulations have been carried out on the computing facilities of the Computational Center for Particle and Astrophysics (C2PAP). Resources supporting this work were provided by the NASA High-End Computing (HEC) Program through the NASA Advanced Supercomputing (NAS) Division at Ames Research Center. Simulations were performed on NASA's Pleiades cluster under award SMD-17-1330. This work used the Extreme Science and Engineering Discovery Environment (XSEDE), which is supported by National Science Foundation grant number ACI-1548562. This work used XSEDE Comet at the San Diego Supercomputer Center (SDSC) through allocation TG-AST170044.

\begin{discussion}

\discuss{Luhmann}{You have assumed very solar-like CME settings. Have you considered that the magnetic fields of the stars may differ, e.g. rapid global field reconfigurations that may not be so confining?}

\discuss{Alvarado-G\'omez}{We have indeed assumed solar CME models for our simulations. While other eruption mechanisms may work in the stellar case, using solar validated models permit a better understanding of the model results with respect to the observations (solar and stellar). This is critical as there are almost no constraints on these type of events in the stellar regime that could inform the models. Regarding the rapid global field configuration, it is important to remember that the typical flare/CME time-scale is much shorter compared to the observed (through spectropolarimetric data) and expected (via dynamo models) evolution of the large-scale field. Therefore, the large-scale confining conditions are expected to be relatively stable, with noticeable changes on activity/magnetic cycle time-scales.}

\end{discussion}




\def\apj{{ApJ}}    
\def\nat{{Nature}}    
\def\jgr{{JGR}}    
\def\apjl{{ApJ Letters}}    
\def\aap{{A\&A}}   
\def\mnras{{MNRAS}}
\let\mnrasl=\mnras
\def\aj{{AJ}}
\def\apss{{AP\&SS}}
\def\planss{{Planetary Space Science}}
\def\solphys{{Solar Physics}}
\def\araa{{Annual Review of Astronomy and Astrophysics}}

\bibliographystyle{aasjournal}
\bibliography{Biblio}

\end{document}